\documentclass[12pt]{article}

\usepackage{latexsym}
\usepackage{amsmath}
\usepackage{amssymb}
\usepackage{bm}
\usepackage{graphicx}
\newcommand{\be}{\begin{equation}}
\newcommand{\en}{\end{equation}}
\newcommand{\eeq}{\end{equation}}
\newcommand{\bear}{\be\begin{array}}
\newcommand{\bea}{\begin{eqnarray}}
\newcommand{\eea}{\end{eqnarray}}
\newcommand{\nn}{\nonumber}
\newcommand{\dd}{\mathrm{d}}
\newcommand{\ee}{\mathrm{e}}
\newcommand{\ii}{\mathrm{i}}
\newcommand{\bk}{{\bf k}}

\newcommand{\bp}{{\bf p}}
\newcommand{\bP}{{\bf P}}
\newcommand{\br}{{\bf r}}

\newcommand{\dst}{\displaystyle}
\newcommand{\fr}[2]{\frac{{\dst #1}}{{\dst #2}}}

\def\lsim{\mathrel{\rlap{\lower4pt\hbox{\hskip1pt$\sim$}}
    \raise1pt\hbox{$<$}}}         %less than or approx. symbol

\def\gsim{\mathrel{\rlap{\lower4pt\hbox{\hskip1pt$\sim$}}
    \raise1pt\hbox{$>$}}}         %greater than or approx. symbol

\date{October 18, 2011}%\today}

\title{Scattering of twisted particles: extension to wave packets
and orbital helicity}
\author{I.~P.~Ivanov$^{1,2}$, V.~G.~Serbo$^{2,3}$
\\
  {\small $^1$ IFPA, Universit\'{e} de Li\`{e}ge, All\'{e}e du 6 Ao\^{u}t 17, b\^{a}timent B5a, 4000 Li\`{e}ge, Belgium}\\
  {\small $^2$ Sobolev Institute of Mathematics, Koptyug avenue 4, 630090, Novosibirsk, Russia}\\
  {\small $^3$ Novosibirsk State University, Pirogova str. 2, 630090, Novosibirsk, Russia}\\
  }

\begin{document}

\maketitle

\begin{abstract}
High-energy photons and other particles carrying non-zero orbital
angular momentum (OAM) emerge as a new tool in high-energy
physics.
Recently, it was suggested to generate high-energy photons
with non-zero OAM (twisted photons) by the Compton backscattering
of laser twisted photons on relativistic electron beams.
Twisted electrons in the intermediate energy range have also been demostrated
experimentally; twisted protons and other particles can in principle be
created in a similar way.
Collisions of energetic twisted states can offer a new look
at particle properties and interactions. A theoretical description
of twisted particle scattering developed previously treated them
as pure Bessel states and ran into difficulty when describing
the OAM of the final twisted particle at non-zero scattering angles.
Here we develop further this formalism by
incorporating two additional important features. First, we treat
the initial OAM state as a wave packet of a finite transverse size
rather than a pure Bessel state. This realistic assumption allows
us to resolve the existing controversy between two theoretical
analyses for non-forward scattering. Second, we describe the
final twisted particle in terms of the orbital helicity --- the OAM projection on its
average direction of propagation rather than on the fixed reaction
axis. Using this formalism, we determine to what extent the
twisted state is transferred from the initial to final OAM
particle in a generic scattering kinematics.
As a particular application, we prove that in the Compton backscattering
the orbital helicity of the final photon stays close to
the OAM projection of the initial photon.

\end{abstract}

\section{Introduction}

\subsection{Particles carrying orbital angular momentum}

In high-energy physics we probe the structure of particles and
their interactions by bringing them into collision and detecting
the products of their scattering. In general, the more control we
have on the initial state particles, the more subtle features we
can measure. For example, sufficiently monochromatic initial beams
allow for a direct measurement of the shape of resonances and of
interference patterns, while a well defined polarization of the
initial particles gives access to spin-dependent structure of
hadrons. It now seems possible that yet another degree of freedom,
the orbital angular momentum (OAM) of the initial states, can be
exploited in the high-energy particle scattering.

Laser beams carrying non-zero OAM are well known in optics,
\cite{OAM}, for a review see \cite{OAMreview},
numerous applications of light with orbital angular momentum are
described in the recent book~\cite{TP}.
The light-field in such a beam is described by a non-plane wave
solution of the Maxwell equations with a helical wave front and an
associated integer winding number $m$. Each photon in this
light-field, which we call a {\em twisted photon}, carries a
non-zero value of OAM projection onto its propagation axis: $L_z =
m\hbar$. An experimental realization~\cite{CKG-2002} exists for
states with OAM projections as large as $m = 200$.

So far, experiments with twisted light were confined mostly to the
optical energy range. However it was recently noted that Compton
backscattering of twisted optical photons off an
ultra-relativistic electron beam can generate high-energy photons
carrying non-zero OAM \cite{serbo1,serbo2}. The technology of
Compton backscattering is well established \cite{backscattering},
thus the realization of this idea seems feasible.

It must be stressed that the possibility to carry OAM is by no mean
an exclusive property of photons. Other particles can carry orbital angular momentum too.
Having a non-zero mass or being a fermion does not forbid the existence
of phase vortices in the transverse plane.
Indeed, following the suggestion made in \cite{bliokh2007},
very recently several groups have reported successful creation of
twisted electrons, first using phase plates
\cite{twisted-electron} and then with computer-generated holograms
\cite{twisted-electron2}. Such electrons carried the energy as
high as 300 keV and the orbital quantum number up to $m\sim 100$.
It is very conceivable that when these electrons are injected into
a linear electron accelerator, their energy can be boosted into
the multi-MeV and even GeV region.
Even more, when beams of protons, neutrons or other particles
with sufficient transverse coherence (and the mere observation
of neutron diffraction on crystals proves that such coherence is achievable)
pass through a specially prepared diffractive grating,
they can gain an OAM as well.

One can therefore imagine that with the future progress in this field
creation of energetic twisted particles of different kind will be possible
and can be used in scattering experiments.
As it was described in \cite{ivanov2011},
the new degree of freedom that enters this scattering process might become
a new promising tool in nuclear and high-energy physics.
For example, it can provide access to such features in the structure of hadrons
which are difficult to probe otherwise.

\subsection{How does OAM change after scattering?}

When a twisted state, be it a photon, an electron or another particle, scatters, elastically or inelastically,
one can ask how its OAM changes after the scattering.
This question received little attention so far, mostly because optical photons with OAM
are almost always assumed to be absorbed rather than scattered.
In the case of twisted electrons, analyses are limited to semiclassical dynamics in external potentials,
see e.g. \cite{bliokh2009}. The fully relativistic, quantum-field-theoretic
treatment of this problem, which is absolutely necessary for high-energy collisions
of twisted particles, has not yet been given.

In fact, this question is also very crucial for the suggestion
of \cite{serbo1,serbo2}.
It was noted that for a strictly backward Compton scattering
the final energetic photon moving along the initial collision axis
carries away exactly the same OAM as the initial photon: $m'=m$.
However this conclusion is valid only for a single point of the
final phase space. In order for this suggestion to become a
reliable technique of generation of high-energy twisted photons
with more or less definite OAM, one must show that $m'\approx m$
holds for small but non-zero angles of the final photons $\beta$,
at least within the range $\beta \lsim m_e/E_e$, where the Compton
scattering receives its dominant contribution to the total cross
section.

This is the point where a controversy in the literature starts. On
one hand, in the original paper \cite{serbo1} it was argued, on
the basis of an approximate consideration of the non-forwards
case, that at very small transverse momentum transfer, $|\bP| \ll
\varkappa$, the final $m'$ stays close to $m$:
 \be
{|m'-m|\over m} \sim {|\bP| \over \varkappa}\,.
 \label{result1}
 \en
Here $P = p'-p$ is the momentum transfer to the electron and
$\varkappa \sim {\cal O}(1\ \mathrm{eV})$ is the conical momentum
spread in the initial twisted state. On the other hand, the
exact non-forward scattering analyzed in \cite{ivanov2011} for a
generic twisted scalar case implies that the entire $m'$-region
contributes homogeneously to the cross section at {\em any}
non-zero transverse momentum transfer:
 \be
m' \in (-\infty,+\infty)\,,\quad \bP \not = 0\,.
 \label{result2}
 \en
These two results are in a clear conflict with each other.

There are two additional reasons to find these results disturbing.
First, if (\ref{result1}) holds, one can expect that for any
reasonable transverse momentum transfer, which is orders of
magnitude larger than $\varkappa$, the final $m'$ should spread
over a very broad range of values with almost no correlation with
$m$. If this were true, that would make the experimental
realization of the suggestion of \cite{serbo1,serbo2} unfeasible.
Second, the result (\ref{result2}) is in sharp contrast with the
conclusion that $m'=m$ for $\bP = 0$, which indicates that there
is no smooth non-forward to forward transition.

In this paper we resolve all these problems by developing further
the formalism of twisted particle scattering. First, we allow the
initial twisted particles to be more or less transversely
collimated wave packets, in contrast to the previously analyzed
case of pure Bessel states. This modification reveals the origin
of the discrepancy between (\ref{result1}) and (\ref{result2}):
they correspond to two different limits in the description of
non-forward to forward transition in twisted state scattering.

Second, following suggestion of \cite{ivanov2011},
we describe each OAM state as a twisted state with
respect to the average propagation axis this very photon rather than
using an OAM projection on an arbitrarily chosen axis.
The OAM projection on the particle averaged propagation direction, which we call the
{\em orbital helicity}, is a more faithful representation of the twisted nature of this state.
This updated formalism leads us to the remarkable
conclusion that for small-angle scattering the quantum number
$m'$ indeed stays close to $m$ even when $|\bP| \gg \varkappa$.

In this paper we consider scalar particle scattering with an
isotropic matrix element.
This simplest set up allows us to focus on the universal
kinematical features of all scattering processes, involving a twisted state of a photon, an electron, a proton etc.
both in the initial and final states.
As a particular example, this includes, but is not limited to,
the Compton scattering of twisted photons.
Our generic scalar analysis is related to these particular cases
just as the scalar theory of diffraction is related to the real diffraction of light
or of matter waves.
In a sense, we investigate what the energy-momentum delta-function
$\delta(p_{f}-p_i)$ turns into when we pass from plane waves to
twisted states. As explained in \cite{ivanov2011}, any non-trivial
matrix element will appear as a multiplicative factor in front of
the resulting expression.

The paper is organized as follows. In Section
\ref{section-description} we introduce the scalar twisted states,
remind the reader how scattering of a twisted particle is
described, and rederive results (\ref{result1}) and
(\ref{result2}). In Section \ref{section-packets} we
introduce the wave packets into description of twisted states
and reconcile these results.
In Section \ref{section-helicity} we generalize the
formalism to include the orbital helicity and finally answer the
question of how twisted the final state is. Section
\ref{section-conclusions} contains our conclusions.

Throughout the paper we use the relativistic units $\hbar = c = 1$.
For a 4-vector $p = (E_p,\vec p)$, we will separate its 3-vector $\vec p$
into the transverse vector $\bp$ and the longitudinal component $p_z.$
Note also that whenever we say forward scattering we actually mean
a scattering at zero transverse momentum transfer, which might be either
strictly forward or strictly backward.

\section{Kinematical features of twisted particle scattering}
\label{section-description}

\subsection{Describing twisted states}

Here we briefly summarize the formalism of Bessel-beam twisted
states introduced in \cite{serbo1}.

We first fix a $z$ axis and solve the free wave equation in
cylindric coordinates $r, \varphi_r, z$. A solution
$|\varkappa,m\rangle$ with definite frequency $\omega$,
longitudinal momentum $k_z$, modulus of the transverse momentum
$|\bk|=\varkappa$ and a definite $z$-projection of orbital angular momentum $m$ has
the form
 \be
|\varkappa, m\rangle = e^{-i\omega t + i k_z z} \cdot
\psi_{\varkappa m}(\br)\,, \quad \psi_{\varkappa m}(\br) = {e^{i m
\varphi_r} \over\sqrt{2\pi}}\sqrt{\varkappa}J_{m}(\varkappa r)\,,
 \label{twisted-coordinate}
  \en
where $J_m(x)$ is the Bessel function.
Such a state is possible for a particle of any mass $M$; the mass appears only in the relation
between $\omega$, $k_z$ and $\varkappa$: $\omega^2 = k_z^2 + \varkappa^2 + M^2$.
The transverse spatial
distribution is normalized according to
 \be
\int d^2\br\, \psi^*_{\varkappa' m'}(\br)\psi_{\varkappa m}(\br) =
\delta_{m m'} \sqrt{\varkappa\varkappa'}\int_0^\infty rdr J_{m}(\varkappa
r) J_{m}(\varkappa' r) = \delta_{m m'}\,
\delta(\varkappa-\varkappa')\,.
 \en
A twisted state can be represented as a superposition of plane
waves:
 \be
|\varkappa,m\rangle = e^{-i\omega t + i k_z z} \int {d^2\bk
\over(2\pi)^2}a_{\varkappa m}(\bk) e^{i\bk \br}\,,
 \label{twisted-def}
  \en
where
 \be
a_{\varkappa m}(\bk)= (-i)^m
e^{im\varphi_k}\sqrt{2\pi}\;{\delta(|\bk|-\varkappa)\over
\sqrt{\varkappa}}\,.
  \en
This expansion can be inverted:
 \be
e^{-i\omega t + i k_z z} \cdot e^{i\bk\br} =  \sqrt{2\pi
\over\varkappa} \sum_{m=-\infty}^{+\infty} i^m e^{-im\varphi_k}
|\varkappa,m\rangle \,,\quad \varkappa = |\bk|\,. \label{PW}
 \en
More details about properties of twisted states, their
normalization and phase space density can be found in
\cite{serbo2,ivanov2011}. Here we just note that although the wave
oscillation amplitude decreases at large radii, the pure Bessel twisted state
of finite amplitude is still not localized and not normalizable in the transverse plane.
Therefore, the intermediate calculations with these Bessel-beam states must be
carried out inside a large but finite cylindric volume of radius
$R$.

\subsection{Collision of twisted particles}

Let us now re-derive the conflicting results (\ref{result1}) and (\ref{result2}).

Consider a $2 \to 2$ scattering of plane wave states with initial
momenta $p$ and $k$ and final momenta $p'$ and $k'$. The
scattering matrix element has the standard form
 \be
S_{PW}(p,k,p',k') = {i (2\pi)^4 \delta^{(4)}(p+k-p'-k')\cdot {\cal M} \over 4\sqrt{E_kE_pE_{k'}E_{p'}}}\,,
 \label{S-planewave}
 \en
where the amplitude ${\cal M}$ is calculated
by Feynman rules. The passage from the plane wave with momentum
$k$ to the twisted state $|\varkappa,m\rangle$ can be performed by
integrating out the plane-wave scattering matrix element over all
$\bk$ with the weight factor $a_{\varkappa m}(\bk)$, as in
(\ref{twisted-def}). If we consider elastic scattering between a
twisted state and a plane wave, we have
 \be
S_{tw} = \int {d^2\bk \over(2\pi)^2} {d^2\bk' \over(2\pi)^2}
a_{\varkappa m}(\bk) a^*_{\varkappa' m'}(\bk')
S_{PW}(p,k,p',k')\,.
 \label{S-twisted}
  \en
Following \cite{serbo1,ivanov2011}, we assume in this and the next
Section that both the initial and the final twisted states here
are defined with respect to the common axis $z$, which is also the
direction of propagation of the initial plane wave particle with
momentum $p$.

\begin{figure}
\centering
\includegraphics[width=0.5\linewidth]{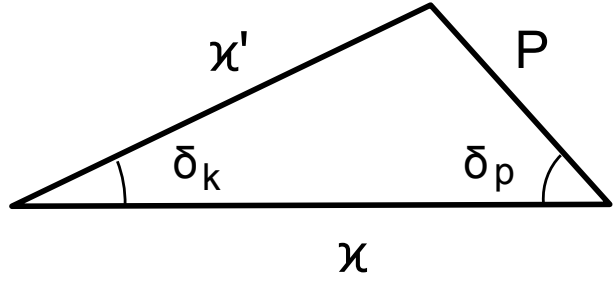}
\caption{\label{fig-triangle} Triangle with the sides $\varkappa$, $\varkappa'$, and $|\bP|$.}
\end{figure}

The integrals in (\ref{S-twisted}) are killed by the transverse
delta-function present in $\delta^{(4)}(p+k-p'-k')$. The
presence of a non-trivial amplitude ${\cal M}$, which is a smooth
function of the momenta, does not influence this integral.
Therefore, the key quantity that enters the twisted scattering
matrix element is the following {\em master integral}
 \bea
&&{\cal I}_{mm'}(\varkappa,\varkappa',\bP) = \int {d^2\bk
\over(2\pi)^2} {d^2\bk' \over(2\pi)^2} a_{\varkappa m}(\bk)
a^*_{\varkappa' m'}(\bk')\cdot (2\pi)^2 \delta^{(2)}(\bk - \bk'
-\bP)\\
 \label{master}
&&=\fr{\ii^{m'-m}}{2\pi}\, \int d^2\bk\; d^2\bk'\, \ee^{\ii
(m\varphi-m'\varphi')}\; \fr{\delta(|\bk|-\varkappa)
\delta(|\bk'|-\varkappa')}{\sqrt{\varkappa \varkappa'}}
\,\delta^{(2)}(\bk - \bk' -\bP) \,.
 \nn
 \eea
where as before $\bP = \bp'-\bp$ is the transverse momentum
transfer, and $\varphi$ and $\varphi'$ are the azimuthal angles of
$\bk$ and $\bk'$, respectively. It is seen from this expression
that the moduli of the transverse momenta satisfy the triangle
rules (Fig.~\ref{fig-triangle}):
 \be
\varkappa \le \varkappa'+|\bP|\,,\quad \varkappa' \le
\varkappa+|\bP|\,,\quad |\bP| \le
\varkappa+\varkappa'\,.
 \label{triangle-rules}
 \en

After integration over ${\bk}$ and $|\bk'|$, the master integral
can be presented in the form
 \be
{\cal I}_{m m'}(\varkappa, \varkappa', \bP)=\fr{A}{2\pi}
\sqrt{\fr{\varkappa'}{\varkappa}} \int_0^{2\pi}\ee^{\ii[
(m-m')\varphi'-m\psi]} \delta\left(\varkappa'
\sqrt{1+\epsilon^2+2\epsilon\cos\varphi'}-\varkappa\right)\dd
\varphi',
 \label{appr1}
 \en
where
 \be
A=\ii^{m'-m}\; e^{i(m-m')\varphi_{P}}\,,\;\;\psi= \arctan
{\fr{\epsilon\sin{\varphi'}}{1+\epsilon\cos{\varphi'}}}\,,\;\;
\epsilon=\fr{|\bP|}{\varkappa'}
 \en
and $\varphi_{P}$ is the azimuthal angle of the momentum transfer $\bP$. The differential cross section is
proportional to $|{\cal I}_{mm'}(\varkappa,\varkappa',\bP)|^2$. It
is this quantity whose $m'$ distribution and
$\varkappa'$-dependence determines the OAM properties of the final
twisted state.

In the case of strictly forward scattering, $\bP=0$,
one immediately obtains
 \be
{\cal I}_{m m'}(\varkappa,\varkappa',0)
=\delta(\varkappa-\varkappa') \delta_{m m'}\,.
 \label{master-forward}
  \en
This result means that in the strictly forward scattering the
twisted quantum numbers $m$ and $\varkappa$ are transferred from
the initial to the final particle without any change,
\cite{serbo1}.

The master integral for the non-zero transverse momentum transfer,
which was calculated in \cite{ivanov2011}, is equal to
 \be
{\cal I}_{m m'}(\varkappa,\varkappa',\bP) =\fr{A}
{2\pi}\fr{\sqrt{\varkappa\varkappa'}}{\Delta} \,\cos[m'\delta_k -
(m-m')\delta_p]\,,
 \label{master-exact}
  \en
where  $\Delta={1\over 2}\varkappa \varkappa' \sin\delta_k$ is
just the area of the triangle with sides $\varkappa$,
$\varkappa'$, $|\bP|$ shown in Fig.~\ref{fig-triangle}, and
 \be
\delta_k = \arccos\left({\varkappa^2+\varkappa^{\prime 2} -
\bP^{2} \over 2 \varkappa \varkappa'}\right)\,, \quad \delta_p =
\arccos\left({\varkappa^2+\bP^{2} - \varkappa^{\prime 2} \over 2
\varkappa |\bP|}\right)
 \label{deltas}
  \en
are its angles.
One sees that the $m'$-dependence of the differential cross
section comes from the oscillating cosine squared:
 \be
|{\cal I}_{mm'}(\varkappa,\varkappa',\bP)|^2 \propto
\cos^2[m'\delta_k - (m-m')\delta_p]\,.
 \en
At very large $m'$ this cosine is a strongly oscillating function
of the moduli of the momenta and it can be replaced by $1/2$.
Therefore, the contribution of arbitrarily large $m'$ is not
suppressed.

As already said above, in an accurate analysis one must keep the
cylindric quantization volume of radius $R$ large but finite. In
this case, the differential cross section receives approximately
homogeneous contribution from the entire $m'$-region \be -m_{max}
\lsim m' \lsim m_{max}\,,\quad m_{max} = \varkappa' R\,. \en
Contribution of each partial wave with a given $m'$ is suppressed
by $1/m_{max}$, so it is the summation
$$
{1 \over m_{max}}\sum_{m'=-m_{max}}^{m_{max}}|{\cal
I}_{mm'}(\varkappa,\varkappa',\bP)|^2
$$
that stays constant in the $R \to \infty$ limit. Therefore, in
this limit we recover the result (\ref{result2}) found in
\cite{ivanov2011}.

On the other hand, one can investigate the small momentum transfer
approximation inside the master integral. Denoting $|\bP| =
\epsilon \varkappa'$ and assuming $\epsilon \ll 1$, one can
represent expression \eqref{appr1} as
 \be
{\cal I}_{mm'}(\varkappa,\varkappa',\bP) \approx \fr{A}{2\pi}
\int_0^{2\pi} e^{\ii[(m-m')\varphi'-m\epsilon\sin\varphi']}
\,\delta(\varkappa'-\varkappa+\varkappa'\epsilon\cos\varphi')\,d\varphi'\,.
 \en
Note that although $\epsilon \ll 1$, the value of $m\epsilon$ in
the exponential can be large. We then perform the formal
small-$\epsilon$ expansion of the delta-functional under the
integral
 \be
\delta(\varkappa'-\varkappa'+\varkappa'\epsilon\cos\varphi') =
\delta(\varkappa'-\varkappa) + \varkappa'\epsilon\cos\varphi' \cdot
\delta'(\varkappa'-\varkappa) + {\it o}(\epsilon)\,,
 \label{functional-expansion}
  \en
which is valid on a class of sufficiently smooth functions of
$\varkappa$ or $\varkappa'$, and keep only the first term. This
leads us to the approximate value for the master integral
 \be
{\cal I}_{mm'}(\varkappa,\varkappa',\bP) =
A\, \delta(\varkappa'-\varkappa) J_{m-m'}(m\epsilon)\,,
 \en
which clearly shows a smooth transition to the strictly forward/backward case
(\ref{master-forward}). From the properties of the Bessel
functions, namely that $J_n(x)$ is strongly suppressed at $x < n$, one
can infer that at small but non-zero transverse momentum transfer
$\bP$ the quantum numbers $m'$ stay close to $m$, and the result
(\ref{result1}) follows, \cite{serbo1}.

\subsection{The origin of the discrepancy}

A careful inspection shows that discrepancy between the results (\ref{result1}) and (\ref{result2})
is linked to the transverse spatial extent of the incoming twisted state.

First of all, as it was already discussed in \cite{ivanov2011},
the discontinuous non-forward to forward transition is a
consequence of the infinite transverse size of the pure
Bessel-beam state. If the radius $R$ of the quantization volume is
kept large but finite, then the smooth transition is restored in
an extremely narrow region of $|\bP| \sim 1/R$. This suggests that
if one replaces the pure twisted initial state
$|\varkappa,m\rangle$ with a wave packet
 \be
|i\rangle = \int_0^{\infty} d\varkappa\, f(\varkappa)
|\varkappa,m\rangle\,,
 \label{wavepacket-def}
  \en
with a narrow weight function $f(\varkappa)$ peaked at
$\varkappa=\varkappa_0$ and having a width $\sigma \ll \varkappa_0$, then
the non-forward to forward transition is expected to be smooth
even for infinite $R$ and to take place within the transverse
momentum transfer region $|\bP| \sim \sigma$.

On the other hand, the derivation of (\ref{result1}) just
reproduced involves manipulation of the delta-functional, which is
valid only if a convolution with a sufficiently smooth function of
$\varkappa$ or $\varkappa'$ is assumed. The pure Bessel states
lack such a convolution, therefore this result cannot be expected
to hold for the pure Bessel states. However, for a sufficiently
compact wave packet this conclusion can be valid.

These observations necessitate a careful re-analysis of the master
integral for a situation when the initial state is described by a
wave packet (\ref{wavepacket-def}) rather than a pure Bessel
state.

\section{Scattering of a twisted wave packet}\label{section-packets}

\subsection{Qualitative features}

Let us start by discussing qualitative features of the
coordinate space wave function of the wave packet
(\ref{wavepacket-def}). When quantitative estimates are needed, we
will use the gaussian approximation for the weight function
$f(\varkappa)$:
 \be
f(\varkappa)=N\exp\left[-\fr{(\varkappa-\varkappa_0)^2}{2\sigma^2}\right]\,,
 \en
with $\sigma \ll \kappa_0$ and the normalization coefficient $N$ fixed by
 \be
\int_0^{\infty} |f(\varkappa)|^2\, \dd \varkappa=1\,.
 \en
The transverse coordinate wave function $\psi_m(r)$ of this wave packet is defined by
 \be
\int_0^{\infty} d\varkappa f(\varkappa)\, \psi_{\varkappa m}(\br)
= {e^{im\varphi_r} \over \sqrt{2\pi}}
\int_0^{\infty}\sqrt{\varkappa}d\varkappa J_m(\varkappa r)
f(\varkappa) \equiv {e^{im\varphi_r} \over \sqrt{2\pi}} \psi_m(r)
 \label{averagedWF}
  \en
and is normalized to unity:
 \be
\int_0^{\infty} rdr\, |\psi_m(r)|^2 = 1\,.
 \en

\begin{figure}
\centering
\includegraphics[width=0.8\linewidth]{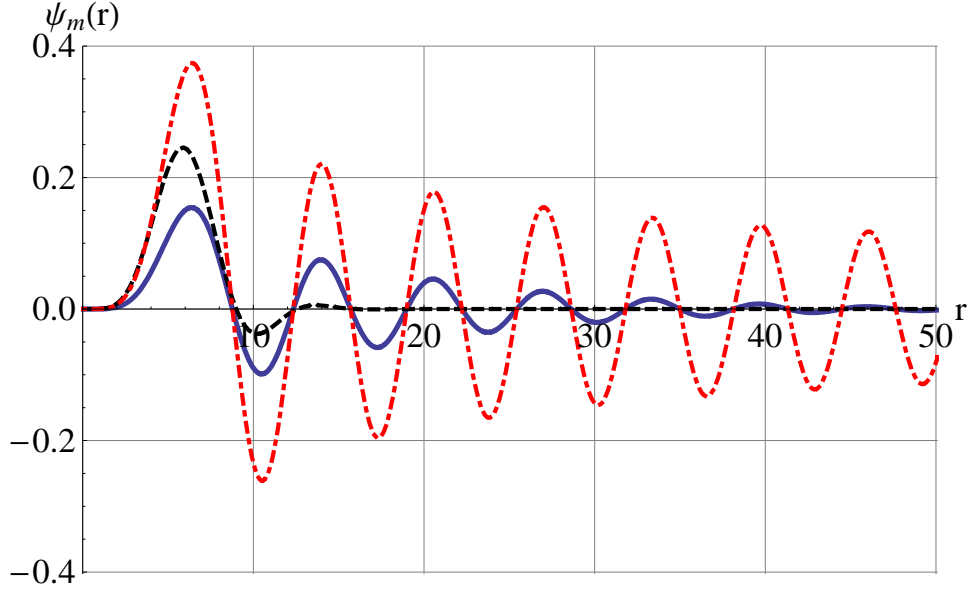}
\caption{\label{fig-radial-wavefunction} Radial wave functions for
$m=5$ and $\varkappa_0=1$: the pure Bessel state
$\sqrt{\varkappa_0}J_5(\varkappa_0 r)$ (dash-dotted line), the
function $\psi_{m}(r)$ for $\sigma=1/20$ (solid line), the
function $\psi_{m}(r)$ for $\sigma=1/5$ (dashed line).}
\end{figure}

In Appendix we study some properties of this averaged wave
function. We show there that if $m$ is not too large, $m \ll
\varkappa_0^2/\sigma^2$, the wave function $\psi_m(r)$ exhibits
radial oscillations characteristic of the Bessel function
$J_m(\varkappa_0 r)$ until $r$ becomes larger than the coherence
radius $r_c = 1/\sigma$. Beyond this radius, the averaged wave
function is strongly suppressed. This behavior is well seen in
Fig.~\ref{fig-radial-wavefunction} where we plotted $\psi_{5}(r)$
for $\sigma=\varkappa_0/5$ and $\sigma=\varkappa_0/20$ for
$\varkappa_0=1$ (in arbitrary units) and compared it with the pure
Bessel state $\sqrt{\varkappa_0}J_5(\varkappa_0 r)$. However, if
$m \gg \varkappa_0^2/\sigma^2$, there is no room left for the
radial oscillations, and the wave function is strongly peaked at
$r = m/\varkappa_0$.

The effective regularization of the radial wave function by the
coherence radius $r_c = 1/\sigma$ plays an important role in the
master integral and its $m'$-dependence. The master integral
(\ref{master}) for the pure twisted states can be also represented
as a triple-Bessel integral, see \cite{ivanov2011}:
 \be
{\cal I}_{m m'}(\varkappa,\varkappa',\bP)  =
A\sqrt{\varkappa\varkappa'} \int_0^\infty rdr\, J_m(\varkappa r)
J_{m'}(\varkappa' r) J_{m-m'}(|\bP| r)\,.
 \label{masterJJJ}
  \en
If $m'$ is extremely large, the main contribution comes from the
large $r$-region, $r \gsim \mathrm{max}(m'/\varkappa', m'/|\bP|)$.
A pure Bessel function for the initial state, $J_m(\varkappa r)$,
is not sufficiently suppressed at such large $r$. However, if the
initial state is a wave packet, then the exponential suppression
is at work beyond $r_c$, which effectively limits the values of
$|m-m'|$.

\subsection{Averaged master integral}

Since the  master integral is a linear functional of the initial
wave function, the averaged master integral can be represented as
 \be
{\cal I}_{mm'}(\varkappa',\bP) = \int_0^{\infty} d\varkappa\,
{\cal I}_{m m'}(\varkappa,\varkappa',\bP) f(\varkappa)
 \label{averaged-master-integral-def}
  \en
or, using Eq.~\eqref{appr1}, as
 \be
{\cal I}_{mm'}(\varkappa',\bP)=\fr{A}{2\pi}
\int_0^{2\pi}\ee^{\ii[ (m-m')\varphi'-m\psi]} \fr{f\left(\varkappa'
\sqrt{1+\epsilon^2+2\epsilon\cos\varphi'}\right)}
{\left(1+\epsilon^2+2\epsilon\cos\varphi'\right)^{1/4}}\dd
\varphi'\,.
 \label{appr1a}
 \en

In this Section we aim at resolving the discrepancy between the
results (\ref{result1}) and (\ref{result2}). Therefore, it is
sufficient to consider non-zero but very small momentum transfer,
$|\bP| = \epsilon\varkappa'$ with $\epsilon \ll 1$. In this case
$\varkappa$ changes in the small interval from $\varkappa'-|\bP|$
to $\varkappa'+|\bP|$ and the averaged master integral can be
approximated as
 \be
{\cal I}_{mm'}(\varkappa',\bP)\approx \fr{A}{2\pi}
\int_0^{2\pi}\ee^{\ii[ (m-m')\varphi'-\epsilon m \sin\varphi']}
\,f\left(\varkappa'+ |\bP|\cos\varphi' \right)\dd \varphi'\,.
 \label{appr1b}
 \en

\begin{figure}
\centering
\includegraphics[width=1.0\linewidth]{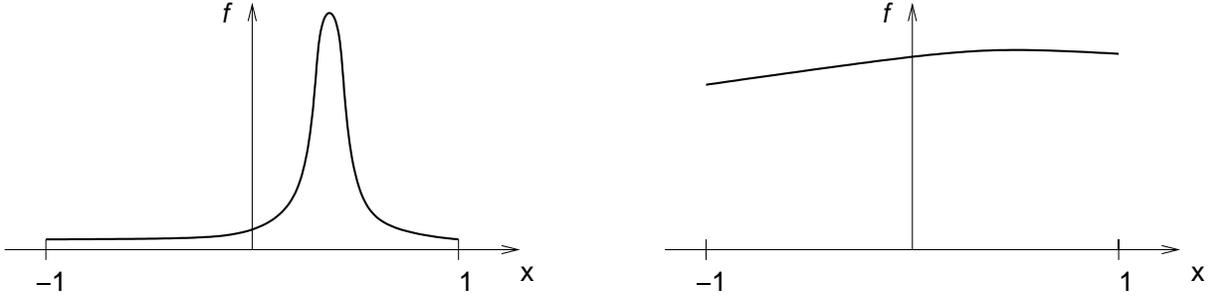}
\caption{\label{two-peaks} The weight function as a function of $x$ in the narrow peak (left) and broad peak (right) regimes.}
\end{figure}

For the further analysis we need to distinguish two cases
depending on the shape of $f(\varkappa'+ |\bP|\cos\varphi')$ as a
function of $ x\equiv \cos\varphi'$ on the interval $[-1,1]$:
 \be
f(\varkappa'+x |\bP|) = N
\exp\left[-{(x-x_0)^2\over2\Delta_x^2}\right]\,,\quad x_0 \equiv
{\varkappa_0-\varkappa' \over |\bP|}\,,\quad \Delta_x \equiv
{\sigma \over |\bP|}\,.
 \en
If it is strongly peaked as in Fig.~\ref{two-peaks}, left, we are
dealing with the narrow-peak situation. It corresponds to $|x_0| <
1$ and $\Delta_x \ll 1$: \be \mbox{narrow peak:}\quad \sigma \ll
|\bP| \ll \varkappa_0\,. \en In this case the relevant integrals
can be taken approximately using Laplace method. The opposite
case, $\Delta_x \gg 1$, corresponds to the broad-peak situation
shown in Fig.~\ref{two-peaks}, right: \be \mbox{broad peak:}\quad
|\bP| \ll \sigma \ll \varkappa_0\,. \en In this case the integrals
can be approximately calculated by using the Taylor
expansion of the weight function.

Let us derive the $m'$ distribution in the broad and narrow peak
situations. In the broad peak case, as the first approximation,
one can put $f\left(\varkappa'+|\bP|\cos{\varphi'}\right)\approx
f\left(\varkappa'\right)$ and immediately obtain the result
 \be
{\cal I}_{mm'}(\varkappa',\bP)\approx A
\,f\left(\varkappa'\right)\, \;J_{m-m'}(\epsilon m)\,.
 \label{appr1c}
 \en
Using the properties of the Bessel functions, one can deduce that
$m'$ should always stay close to $m$, otherwise the contribution
is suppressed. For small values of $m$, when $m\epsilon \ll 1$,
the only significant contribution comes from $m'=m$. For large
$m$, when $m\epsilon \gg 1$, the $m'$-distribution is spread over
several values around $m' \approx m$.

For the narrow peak case, $|m-m'| \gg m\epsilon$ still holds
almost in the entire $m'$ region. Then, we can read off (\ref{appr1b})
that the averaged master integral effectively extracts the $(m-m')$-th
Fourier harmonic of the weight function. Since the weight
function has a peak with a width $\Delta_x$, we conclude that
 \be
|m-m'| \lsim {1 \over \Delta_x} = {|\bP| \over
\sigma}\,.
 \label{m'narrow}
  \en
This means that the final $m'$ can strongly differ from and be
much larger than the initial $m$ if $\sigma$ is sufficiently small.
Considering additionally the
region $|m-m'| \lsim m\epsilon$ does not change this conclusion.

\subsection{Different limits}

The key quantity which we discuss in this Section is the
$m'$-distribution of the scattering matrix element at small $\bP$.
Our analysis reveals that the answer to this question depends in fact
on a subtle interplay between two different limits:
$\sigma \to 0$ and $|\bP| \to 0$. The apparently conflicting
results (\ref{result1}) and (\ref{result2}) simply correspond to
two different choices of which limit is taken first.

If the initial wave packet is fixed (i.e. $\sigma$ is kept
constant) and $|\bP| \to 0$, then we are in the broad-peak regime,
and the result (\ref{result1}) follows. In this way we study the
non-forward to forward transition for a wave packet of fixed size.
This implicit assumption, although not mentioned in \cite{serbo1},
is essential when deriving this result.

On the contrary, if $|\bP|$ is kept constant but $\sigma$
decreases, we enter the narrow-peak regime, and (\ref{m'narrow})
holds. In the $\sigma \to 0$ limit we recover the result
(\ref{result2}). This limit corresponds to the point-by-point
analysis of the small-$|\bP|$ behaviour in the scattering of true
Bessel states, studied in \cite{ivanov2011}.
This resolves the discrepancy mentioned in the introduction.

\section{Orbital helicity of the final twisted state}
\label{section-helicity}

In the previous Section we reconciled results (\ref{result1})
and (\ref{result2})
which was a technical rather than physical problem. However our
calculations did not answer the physically important question:
what are the true OAM properties of the final twisted state with
respect to {\em its own} propagation direction?

The results of the previous Section imply that if the momentum transfer $|\bP|\gg
\varkappa$, a very broad $m'$-region contributes to the
differential cross section. However, this broad region can easily
be an artefact of using the same $z$ axis to describe both the initial and the final
twisted states. Indeed, (\ref{PW}) shows that a non-forward plane
wave, when expanded in the basis of twisted states with respect to
axis $z$, involves all values of $m$ from minus to plus infinity,
despite the fact that it actually carries no OAM at all. As it
was suggested in \cite{ivanov2011}, for the full resolution of
this problem one needs to introduce the concept of {\em orbital
helicity}: the projection of orbital angular momentum not on the
reaction axis $z$ but on the axis of the average propagation
direction of the final particle. This is what we do in the present
Section.

\subsection{Orbital helicity distribution: pure Bessel states}

Let us consider again the scattering process \be
|\varkappa,m\rangle_{\vec n} + |PW(\vec p)\rangle \to
|\varkappa',m'\rangle_{\vec n'} + |PW(\vec p')\rangle
\label{off-forward} \en with fixed $\varkappa$, $(\vec k \vec n)
\equiv k_z$, $m$ and fixed 3-momentum transfer $\vec P \equiv \vec
p'-\vec p$. The twisted states in (\ref{off-forward}) are defined
with respect to two different axes. The incoming state is written
with respect to the direction $\vec n$ which coincides with the
axis $z$. The final twisted state is defined with respect to
another direction $\vec n' \not = \vec n$ to be defined below.
The average value of
the 3-momentum in the initial and final states are
\be
\langle \vec k \rangle \equiv \langle \varkappa,m| \vec k
|\varkappa,m\rangle = k_z \vec n\,,\quad
\langle \vec k' \rangle \equiv \langle \varkappa',m'| \vec k' |\varkappa',m'\rangle
= k'_{z'} \vec n'\,. \en Note that $k'_{z'}$ means here the component of the
vector $\vec k'$ along the axis $z'$ directed along $\vec n'$.

{\em A priori}, any choice of the axis $\vec n'$ is allowed. We find
it convenient to use the following prescription: we draw $\vec n'$
along the direction of the well-defined 3-momentum $\vec q \equiv
\langle \vec k\rangle - \vec P \equiv q\vec n'$. Note that
although $\vec q$ and $\langle \vec k' \rangle$ are parallel, they
are not equal, $q \not = k'_{z'}$, because the averaged momenta are
not supposed to obey any conservation law \be \langle \vec
k\rangle \not = \langle \vec k'\rangle + \vec P\,. \en In this way
the quantity $m'$ becomes the {\em orbital helicity} rather than
just OAM projection on an arbitrary axis.

We now aim at evaluating the master integral for this scattering.
When calculating the scattering matrix element for the process
(\ref{off-forward}), we have a very familiar expression for the
master integral:
 \bea
{\tilde {\cal I}}_{mm'} &\equiv& 2\pi\, \delta(k_z - k'_z - P_z) \cdot
{\cal I}_{mm'}(\kappa,\kappa',\bP)\nonumber\\
&=& i^{m'-m} \sqrt{\varkappa\varkappa'}\int_0^{2\pi} d\varphi
d\varphi'\, e^{i(m\varphi - m'\varphi')}\, \cdot
\delta^{(3)}(\vec k - \vec k' - \vec P)\,.
 \label{master1}
  \eea
Notice two important differences with respect to the ``coaxial''
master integral (\ref{master}). First, the azimuthal angles
$\varphi$ and $\varphi'$ lie in two different planes
which are orthogonal to $\vec n$ and $\vec n'$, respectively.
Second, the master integral contains 3-dimensional delta-function
instead of just two-dimensional: since the integration is not limited to a single
transverse plane, the longitudinal delta-function is,
naturally, included in the integral.

\begin{figure}
\centering
\includegraphics[width=0.8\linewidth]{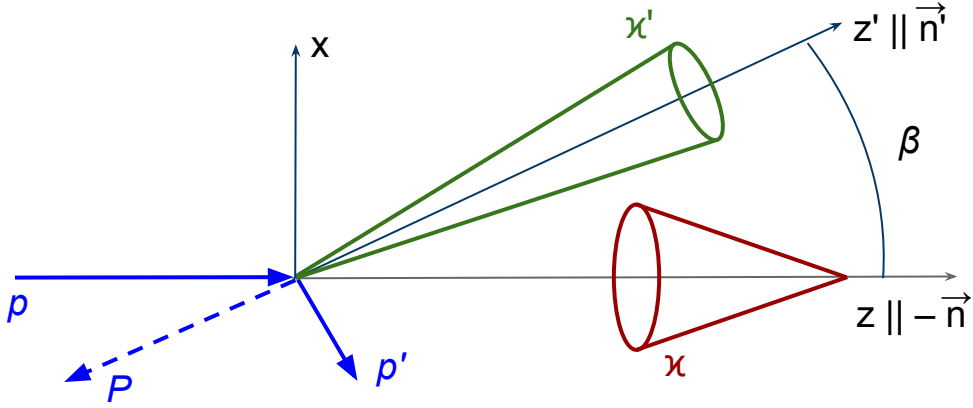}
\caption{\label{off-forward-fig} Kinematics and choice of the axes for the off-forward twisted particle scattering.}
\end{figure}

The integral (\ref{master1}) can be easily done in the usual
cartesian coordinate frame $(x,y,z)$. This frame is fixed in the
following way (see Fig.~\ref{off-forward-fig}). The directions
$\vec n$ and $\vec n'$ define the axis $z$ and the $(x,z)$ plane.
The axis $y$ is orthogonal to the plane containing $\vec n$, $\vec
n'$, $\vec P$.
In this coordinate frame the momenta $\vec k$ and $\vec P$ are
 \be
\vec k = \left(\begin{array}{c} \varkappa  \cos\varphi \\
\varkappa \sin\varphi \\ k_z \end{array}\right)\,,\quad \vec P =
\left(\begin{array}{c} P_x \\ 0 \\ P_z \end{array}\right)\,,
 \en
with $P_x < 0$. The direction of $\vec n'$ is
$(\sin\beta,0,\cos\beta)$, which is defined by
\be
\vec q \equiv \langle \vec k\rangle - \vec P = \left(\begin{array}{c} -P_x \\ 0 \\
k_z-P_z \end{array}\right) = q \left(\begin{array}{c} \sin\beta
\\ 0 \\ \cos\beta\end{array}\right)\,, \en where $q^2 = P_x^2 +
(k_z - P_z)^2$. Then, the momentum $\vec k'$ has components \be
\vec k' = \left(\begin{array}{c}
\varkappa'c_\beta c_{\varphi'} + k'_{z'} s_\beta \\
\varkappa's_{\varphi'} \\
-\varkappa's_\beta c_{\varphi'} + k'_{z'} c_\beta
\end{array}\right)\,,
\en
where we introduced the obvious
short-hand notation for sines and cosines.

The three-dimensional delta-function in (\ref{master1}) expresses
the conservation of the three-momentum at the level of plane
waves: $\vec k = \vec k' + \vec P$. Writing it explicitly, we
obtain
\bea
 \delta^{(3)}(\vec k - \vec k' - \vec P) &=&
\delta(\varkappa c_\varphi - \varkappa' c_\beta c_{\varphi'} -
(k'_{z'}-q)s_\beta)\nonumber\\
& \times & \delta(\varkappa s_\varphi-\varkappa's_{\varphi'})
\cdot \delta(\varkappa' s_\beta c_{\varphi'} - (k'_{z'}-q)c_\beta)\,.
\eea
In order for the integral to be non-zero, we require that \be
\varkappa' s_\beta \ge |k'_{z'}-q| c_\beta\,.\label{condition1} \en
Then the $z$-delta-function can be used to kill the $\varphi'$
integration. Let us for the moment drop the
$\exp(im\varphi-im'\varphi')$ factor from the master integral
(\ref{master1}). Then we have \bea
&& \int d\varphi d\varphi'  \delta^{(3)}(\vec k - \vec k' - \vec P) \nonumber\\
&=& {2 \over s_\beta} \int d\varphi\,
\delta(\varkappa^{\prime 2}s^2_{\varphi'}-\varkappa^2 s^2_\varphi)\ \delta\!\left(\varkappa c_\varphi - {k'_{z'}-q \over s_\beta} \right)\nonumber\\
&=& {2 \over s_\beta}\ \delta\!\left[\varkappa^{\prime 2} + (k'_{z'}-q)^2 - \varkappa^2\right] \int d\varphi \ \delta\!\left(\varkappa c_\varphi - {k'_{z'}-q \over s_\beta} \right)\nonumber\\
&=& {2 \over \sqrt{\varkappa^{\prime 2}-\varkappa^2\cos^2\beta}}\
\delta\!\left[\varkappa^{\prime 2} + (k'_{z'}-q)^2 -
\varkappa^2\right]\,.\label{master2}
\eea
 Here we implicitly
assumed that the integration over $\varphi$ goes not around the
full $2\pi$ domain, but only over the semicircle where
$\sin\varphi$ has the same sign as $\sin\varphi'$.
The extra delta-function in (\ref{master2}) replaces
$\delta(k_z-k'_z-P_z)$ in the ``coaxial'' case. Here it leads to
important conclusions that
\be
\varkappa' \le \varkappa\,,\quad |k'_{z'} - q| \le \varkappa\,.
\en
So, the values of the momentum
transfer ${\vec P}$ (both $P_x$ and $P_z$) can be large,
$|P_{x,z}| \gg \varkappa$, but they must be accompanied by a
correspondingly large value of $k'_{z'}$.

It is convenient to introduce the ``angle'' $\xi$ by
\be
\varkappa' \equiv \varkappa \cos\xi\,,\quad k'_{z'}-q =
\varkappa\sin\xi\,,\quad \xi \in [-\pi/2,\pi/2]\,.
\label{angle-xi} \en In this notation condition (\ref{condition1})
becomes \be \sin(\beta-|\xi|) \ge 0 \quad \to \quad \beta \ge
|\xi|\,.\label{condition1a} \en which also implies $\varkappa^2
\ge \varkappa^{\prime 2} \ge \varkappa^2\cos^2\beta$. The pair of
values of $\varphi$ and $\varphi'$, which are set by the
delta-functions, can be written as
\be
\cos\varphi = {\sin\xi
\over \sin\beta}\,,\quad \cos\varphi' = {\tan\xi  \over \tan\beta}\,.\label{anglesphiphi'}
\en

Let us now recall that the master integral (\ref{master1})
contains the exponential factor $\exp(im\varphi-im'\varphi')$.
Since the integral just taken receives its contribution only from
two $(\varphi,\varphi')$ points, this extra factor is simply an
overall multiplier computed at each of these two points.
Effectively, it corresponds to the replacement $2 \to
2\cos(m\varphi-m'\varphi')$. Thus, the final result for the master
integral is
 \be
{\tilde {\cal I}}_{mm'} = i^{m'-m} \sqrt{\varkappa\varkappa'}
\cdot 2{\cos(m\varphi-m'\varphi') \over  \sqrt{\varkappa^{\prime
2}-\varkappa^2\cos^2\beta}} \ \delta\!\left[\varkappa^{\prime 2} +
(k'_{z'}-q)^2 - \varkappa^2\right]\,,
 \label{master3}
  \en
where the values of $\varphi$ and $\varphi'$ are given by
(\ref{anglesphiphi'}). This expression shows the final {\em
orbital helicity} $m'$ distribution. Similarly to the ``coaxial''
case, ${\tilde {\cal I}}_{mm'}$ is an oscillatory function of
$m'$. Clearly, $|\tilde{\cal I}_{mm'}(\varkappa, \varkappa')|^2$
is not suppressed even at extremely large $m'$. However, this is
an artefact of taking pure Bessel beams for the initial and final
twisted states. In reality, a pure Bessel state is as unphysical
as a plane wave, because its radial coordinate wave function
effectively extends to infinity and is not normalizable. Even if
$m$ is small, all values of $m'$ up to infinity contribute to the
cross section.

\subsection{Orbital helicity distribution: wave packets}

The above conclusion is expected to change if the initial and
final twisted states are assumed to be a wave packets rather than
pure Bessel states. As before, we represent the initial and final
twisted states similarly to (\ref{averagedWF}): initial twisted
state as a superposition of $|\varkappa, m \rangle$ states with
equal value of $m$ and different values of $\varkappa$
(distributed by the weight function $f(\varkappa)$ around
$\varkappa_0$ in a region with width $\sigma$) and final twisted
state as a superposition of $|\varkappa', m' \rangle$ states with
equal value of $m'$ and different values of $\varkappa'$
(distributed by the weight function $g(\varkappa')$ around
$\varkappa'_0$ in a region with width $\sigma'$).

Then we calculate the averaged master integral
 \be
{\tilde {\cal I}}_{mm'}^{av}= \int d \varkappa f(\varkappa) \int d
\varkappa' g(\varkappa')\,{\tilde {\cal I}}_{mm'} (\varkappa,
\varkappa')\,.
 \en
Since the initial laser photon is assumed to be monocromatic, the
final photon has a fixed energy, $\omega'=E+\omega-E'$, therefore,
the longitudinal momenta
 \be
k_z=\sqrt{\omega^2-\varkappa^2}\,,\;\;
k'_{z'}=\sqrt{(\omega')^2-(\varkappa')^2}
 \label{kz-kz}
 \en
change when $\varkappa$ and $\varkappa'$ are varied.

Let us first use the $\varkappa'$ integration to eliminate the
delta-function. Manipulation with delta-function gives:
 $$
\delta\!\left[\varkappa^{\prime 2} + (k'_{z'}-q)^2 -
\varkappa^2\right]=\delta\left(\varkappa^2-\omega'^2
-q^2+2q\sqrt{\omega'^2-\varkappa'^2} \right)=
\fr{q^2+\omega'^2-\varkappa^2}{8qD}\,\delta\left(\varkappa'-
\fr{2D}{q} \right)\,,
 $$
where
 $$
D= \fr 14 \sqrt{2\left(\varkappa^2\omega'^2+\omega'^2q^3+
q^2\varkappa^2\right) -\varkappa^4-\omega'^4-q^4}\,.
 $$
After the $\varkappa'$ integration we obtain
 \be
{\tilde {\cal I}}_{mm'}^{av}=\int d \varkappa f(\varkappa)
g(\varkappa')\,i^{m'-m} \sqrt{\fr{\varkappa}{\varkappa'}} \cdot
2{\cos(m\varphi-m'\varphi') \over  \sqrt{\varkappa^{\prime
2}-\varkappa^2\cos^2\beta}} \;
\fr{q^2+\omega'^2-\varkappa^2}{4q^2}
 \en
with $\varkappa'= 2D/q$.

The resulting expression for ${\tilde {\cal I}}^{av}_{mm'}$ can be
easily analyzed in the case of small-angle scattering, $\beta \ll
1$, which implies $|\xi|\leq \beta \ll 1$ and $\varphi \approx
\varphi'$. As the result, the cosine becomes simply
 \be
{\cos[(m-m')\arccos(\xi/\beta)] \over \sqrt{\cos^2\xi -
\cos^2\beta}} = {T_{m'-m}(\xi/\beta) \over \sqrt{\beta^2-\xi^2}},
 \en
where $T_n(x)$ is the Chebyshev's polynomial of the first kind:
$$
\cos(n\cdot \arccos(x)) = T_n(x)\,.
$$
The integral is then proportional to
 \be
\int d \varkappa f(\varkappa)
g(\varkappa')\;\fr{T_{m'-m}(\xi/\beta)} {\sqrt{\beta^2-\xi^2}}=
\int d\xi\, F(\xi)\; \fr{T_{m'-m}(\xi/\beta)}
{\sqrt{\beta^2-\xi^2}}\,,
 \en
where $F(\xi)$ includes both weight functions and the Jacobian of
the $\varkappa \to \xi$ change of variables.

This Chebyshev polynomial makes $(m'-m)/4$ oscillations on the
interval $\xi/\beta$ from $-1$ to $1$. Smearing of $\varkappa$ due
to the weight function leads to smearing of $\xi/\beta$ by the
amount of $\Delta \xi/(8\beta)$. Therefore, if $|m'-m| \gg
8\beta/\Delta\xi$, the oscillations strongly suppress the
contribution. We conclude that only $m'$ such that
 \be
|m'-m| \lsim \fr{8\beta}{\Delta \xi}
  \en
effectively contribute to the cross section.

Let us estimate $\Delta \xi$ using expression
 \be
\xi \approx \fr{k'_{z'}-q}{\varkappa}
 \en
obtained from Eq.~\eqref{angle-xi}. From here we have
 \be
\Delta\xi \approx \fr{\Delta k'_{z'}-\Delta q}{\varkappa}-
\fr{k'_{z'}-q}{\varkappa^2}\,\Delta\varkappa \approx -\fr{\Delta
q}{\varkappa}\approx \fr{\Delta\varkappa}{\omega}\approx
\fr{\sigma}{\omega}\,,
 \en
since the following estimations are valid from~\eqref{kz-kz} at
$\omega\ll \omega'$ and $\varkappa'\approx \varkappa$:
 \be
\Delta k_z\approx -\fr{\varkappa\Delta\varkappa}{\omega}\,,\;\;
\Delta k'_{z'}\approx
-\fr{\varkappa'\Delta\varkappa'}{\omega'}\ll\Delta k_z\,,
 \en
and therefore,
 \be
\Delta q=\Delta \sqrt{(k_z-P_z)^2+P_x^2}=\fr{(k_z-P_z)\Delta
k_z}{q}\approx -\fr{\varkappa}{\omega}\,\Delta \varkappa\,.
 \en
As a result, we obtain the important estimate:
 \be
|m'-m| \lsim 8\,\beta\;\fr{\omega}{\sigma}\,.
  \label{true-m'}
 \en
Note that the right hand side of this inequality can be small for
small enough angles $\beta$.

A crucial observation is that the result (\ref{true-m'}) does {\em
not} require the transverse momentum transfer $|\bP|$ to be small
(smaller than $\varkappa_0$ or even $\sigma$ as it was the case in
the previous Section). The only assumption for this simple
analysis was that the scattering angle was much smaller than one.
We remind once again that for the Compton scattering, where the
region of small scattering angles $\beta\sim m_e/E_e$ gives the
dominant contribution to the total cross section, this
approximation is fully valid.

Therefore, we proved that if the initial and final twisted states
are the wave packets, then final orbital helicity $m'$ stays close
to $m$ and the final $\varkappa'$ stays close to $\varkappa$
during small-angle scattering.

\section{Conclusions}\label{section-conclusions}

Photons carrying non-zero orbital angular momentum are now
routinely produced in the low-energy domain. It was recently
argued in \cite{serbo1,serbo2} that Compton backscattering of
photons with OAM can help create high-energy photons carrying
large values of OAM.
Intermediate energy electrons with large OAM have been recently created experimentally,
and the same technique can in principle be used
to give twist to other particles, such as protons.
Thus, energetic particles with OAM and their scattering
can become a new promising tool in
experimental subatomic physics.

In these circumstances, it is essential to understand quantum-field-theoretically
how the OAM of a particles changes after its scattering.
In particular, the vital question about the OAM of the final twisted photon
scattered at non-zero angle, which arises in the context of Compton backscattering,
has not been satisfactorily answered so far.
Two previous analyzes, \cite{serbo1} and \cite{ivanov2011},
led to conflicting results. In order to resolve this controversy and to answer
the question, one needs to generalize the formalism and to allow the initial twisted state
to be in the form of a wave packet rather than
a pure Bessel state.
This generalization of the formalism is done in the present paper.
Using it, we completely resolve the existing controversy by showing that
the two apparently conflicting results emerge in two different limits of the same expression.

Another key part of this work is the concept of orbital helicity of a twisted particle:
that is, the OAM projection on its averaged propagation direction
rather than on the reaction axis.
With this more physically appealing definition
we derived the orbital helicity distribution of a generic scalar scattering at any angle.
Our results confirm the intuitive expectation that the orbital helicity
is approximately conserved at small-angle scattering.

\section*{Acknowledgements}
The authors thank I.~Ginzburg for valuable comments. This work was
supported by the Belgian Fund F.R.S.-FNRS via the contract of
Charg\'e de recherches, and in part by grants of the Russian
Foundation for Basic Research 09-02-00263-a and 11-02-00242-a
as well as NSh-3810.2010.2.

\appendix

\section{Wave packets of Bessel states}

Here, for the sake of completeness, we study the radial wave
function (\ref{averagedWF}) of the Bessel-beam wave packet in some
detail.

Let us start with its expected qualitative behaviour if $m$ is not
too large. As $r$ grows, the radial oscillations of different
twisted states in the wave packed remain almost in phase up to the
{\em coherence radius} $r_c = 1/\sigma$, and at $r> r_c$
destructive interference strongly suppresses the wave
function. However, at extremely large radii the true large-$r$
asymptotics determined by the small-$\varkappa$ behavior of the
weight function sets in.

The calculations below show that this qualitative picture changes
when $m \gsim \varkappa_0^2/\sigma^2$. Depending on whether $m$ is
larger or smaller than $ \varkappa_0^2/\sigma^2$, we will speak of
the small-$m$ and large-$m$ regions.

To study the shape of $\psi_m(r)$, we break it into three
$r$-regions: the central hole, $\varkappa_0 r <m$, the
intermediate region, $\varkappa_0 r \gg m$, and the true large-$r$
asymptotics, which sets in at $r > r_*$ to be determined below.
In the first two regions, the approximate forms of the Bessel
functions allow us to represent $\psi_m(r)$ as \be \psi_m(r)
\approx N  \sqrt{2\pi} \sigma \sqrt{\varkappa_0} J_{m}(\varkappa_0
r) \times \left\{
\begin{array}{lcl}
\exp\left[-\fr{\sigma^2 (m+1/2)^2}{2\varkappa_0^2}\right] &\mbox{at} & \varkappa_0 r \lsim m \,,\\[4mm]
\exp\left[-\fr{\sigma^2 r^2}{2}\right] &\mbox{at} & \varkappa_0 r
\gg m\,.
\end{array}
\right.
 \label{regions12}
 \en
These shapes come from the region under the peak in  the weight function $f(\varkappa)$
which is assumed to be its dominating feature.

The very far asymptotics of $\psi_m(r)$ comes from the
small-$\varkappa$ behaviour of $f(\varkappa)$. For example, if
$f(0)$ is finite, then $f(0)$ can be taken out of the integral,
and we are left with an integral of the form of \be I(y)=\int_0^y
dx \sqrt{x} J_m(x)\,. \en The large $y$ behavior of this intregral
is
$$
I(y) \approx \sqrt{2}{\Gamma\left({3\over 4} + {m \over 2}\right) \over \Gamma\left({1\over 4} + {m \over 2}\right)}
- \sqrt{{2 \over \pi}}\cos\left(y - {\pi \over 2}m + {\pi \over 4}\right) + {\cal O}(y^{-1})\,.
$$
The first term is very close to $\sqrt{m}$ at all $m \ge 1$, while
the second term is a small amplitude oscillatory function. Thus, a
reasonable estimate for the large-$r$ asymptotics of the wave
function is
\be
\psi_m(r) \approx N\, \exp\left[-{\varkappa_0^2 \over
2\sigma^2}\right]\cdot {\sqrt{m} \over r^{3/2}}\,. \label{region3}
\en
where the exponentially suppressed factor is simply $f(0)$.
Note that the exact $r$-power is driven by the $f(\varkappa)$
behaviour in the $\varkappa \to 0$ limit; the asymptotics
$\psi_m(r) \propto r^{-3/2}$ is a consequence of the finite
$f(0)$.

By comparing (\ref{regions12}) and (\ref{region3}), one finds that
the true large-$r$ asymptotics sets in beyond
 \be
r_* = {\varkappa_0 \over \sigma^2} \gg r_c = {1 \over \sigma}\,.
 \en
The intermediate region exists if $r_* \gg m/\varkappa_0$.
Therefore, the qualitative picture described above holds if $m \gg
\varkappa_0^2/\sigma^2$, that is, in the small-$m$ region.

For very large $m$, there is no room for the intermediate
$r$-region. The radius at which the Bessel function $J_m$ is
supposed to have its first peak is already so large, that
destructive interference takes place. Therefore, the wave function
exhibits no radial oscillations and has the shape of a single peak
at $r = m/\varkappa_0$. Its shape can be estimated to be
 \be
\psi_m(r) \propto {\sqrt{\varkappa_0}\over
r}\exp\left[-{(m-\varkappa_0 r)^2 \over 2 r^2 \sigma^2}\right]
\approx {\varkappa_0^{3/2} \over m}\exp\left[-{(r-m/\varkappa_0)^2
\over 2 \Delta^2}\right] \,,
 \en
with $\Delta = m\sigma/\varkappa_0^2$ (so that $\Delta/r =
\sigma/\varkappa_0$).

\end{document}